\newcommand\apj{{ApJ\,}}%
\newcommand\apjl{{ApJ\,}}%
\newcommand\apjs{{ApJS\,}}%
\newcommand\aap{{A\&A\,}}%
\newcommand\mnras{{MNRAS\,}}%
\newcommand\nat{{Nature\,}}%
\newcommand\msun{{$M_\odot$}}

\newcommand\rsun{{$R_\odot$}}

\newcommand{\Kepler}{\textit{Kepler}}

\documentclass[graybox, natbib, footinfo]{svmult}

\usepackage{natbib}       
\usepackage{journals}

\usepackage{mathptmx}       
\usepackage{helvet}         
\usepackage{courier}        
\usepackage{type1cm}        
\usepackage{makeidx}         
\usepackage{graphicx}        
\usepackage{sidecap}
\usepackage{multicol}        
\usepackage[bottom]{footmisc}
\usepackage{url}

\makeindex             

\begin{document}

\title*{Adiabatic solar-like oscillations in red giant stars}
\author{Josefina Montalb\'an \and Andrea Miglio \and Arlette Noels \and Richard Scuflaire \and Paolo Ventura \and Francesca D'Antona}
\authorrunning{J.~Montalb\'an et al.}

\institute{J.~Montalb\'an \and A.~Noels \and R.~Scuflaire
            \at 
            Institut d'Astrophysique et Geophysique, Universit\'e de Li\`ege,
            all\'ee du 6 Ao\^ut 17, B-4000 Li\`ege, Belgium\\
           \email{j.montalban, a.noels, r.scuflaire@ulg.ac.be}
	    \and
	   A. Miglio \at School of Physics and Astronomy, University of Birmingham, Edgbaston, Birmingham B15 2TT, United Kingdom\\
              \email{miglioa@bison.ph.bham.ac.uk}
	    \and
	   P. Ventura \and F. D'Antona \at INAF-Roma, Via Frascati,33, Monteporzio Catone, Rome, Italy\\
             \email{paolo.ventura, franca.dantona@oa-roma.inaf.it}  
}

%
%
\maketitle

\abstract{Since the detection of non-radial solar-like oscillation modes in red giants with the CoRoT satellite, the interest in the asteroseismic properties of red giants and the link with their global properties and internal structure is substantially increasing. Moreover, more and more precise data are being collected with the space-based telescopes CoRoT and {\it Kepler}. In this paper we present a survey of the most relevant  theoretical and observational results obtained up to now concerning the potential of solar-like oscillations in red giants.}

\section{Structure and oscillation spectra of red giant models}
\label{sec:1}

\begin{figure}[b]
\includegraphics[scale=.025]{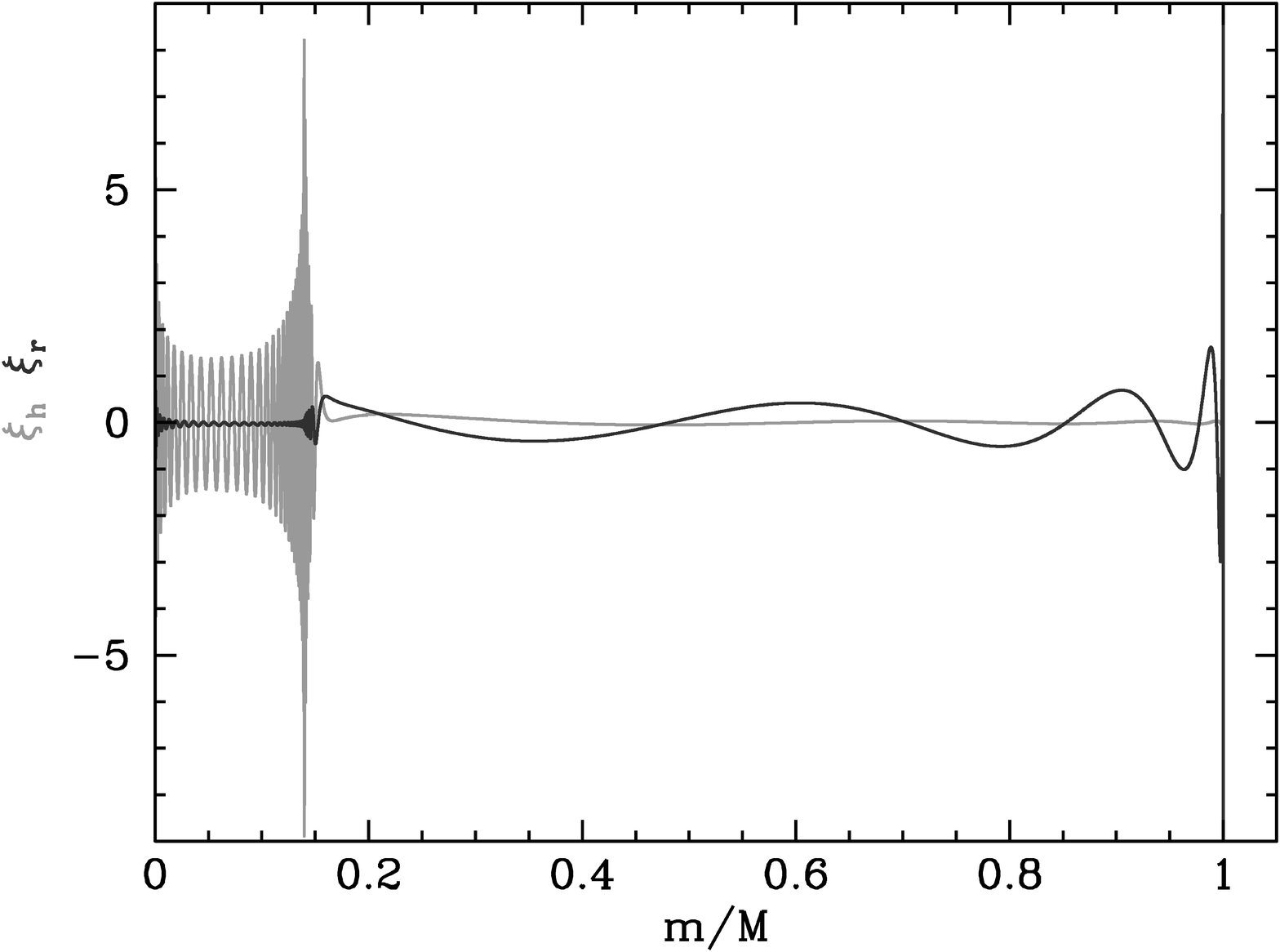}\hfill
\begin{minipage}[t]{0.33\linewidth}
\vspace*{-4.5cm}
\caption{Horizontal (grey) and radial (black) components of displacement corresponding to a mixed mode as a function of the relative mass for a 1.4~\msun\ model with 5~\rsun. g modes are predominantly horizontal while acoustic ones are radial. This mixed mode behaves as a g mode in the central region ($m/M < 0.15$) and as a p mode in the envelope.}
\label{fig_mixmod}       
\end{minipage}\hfill

\end{figure}

\begin{figure}[b]
\includegraphics[scale=.0335]{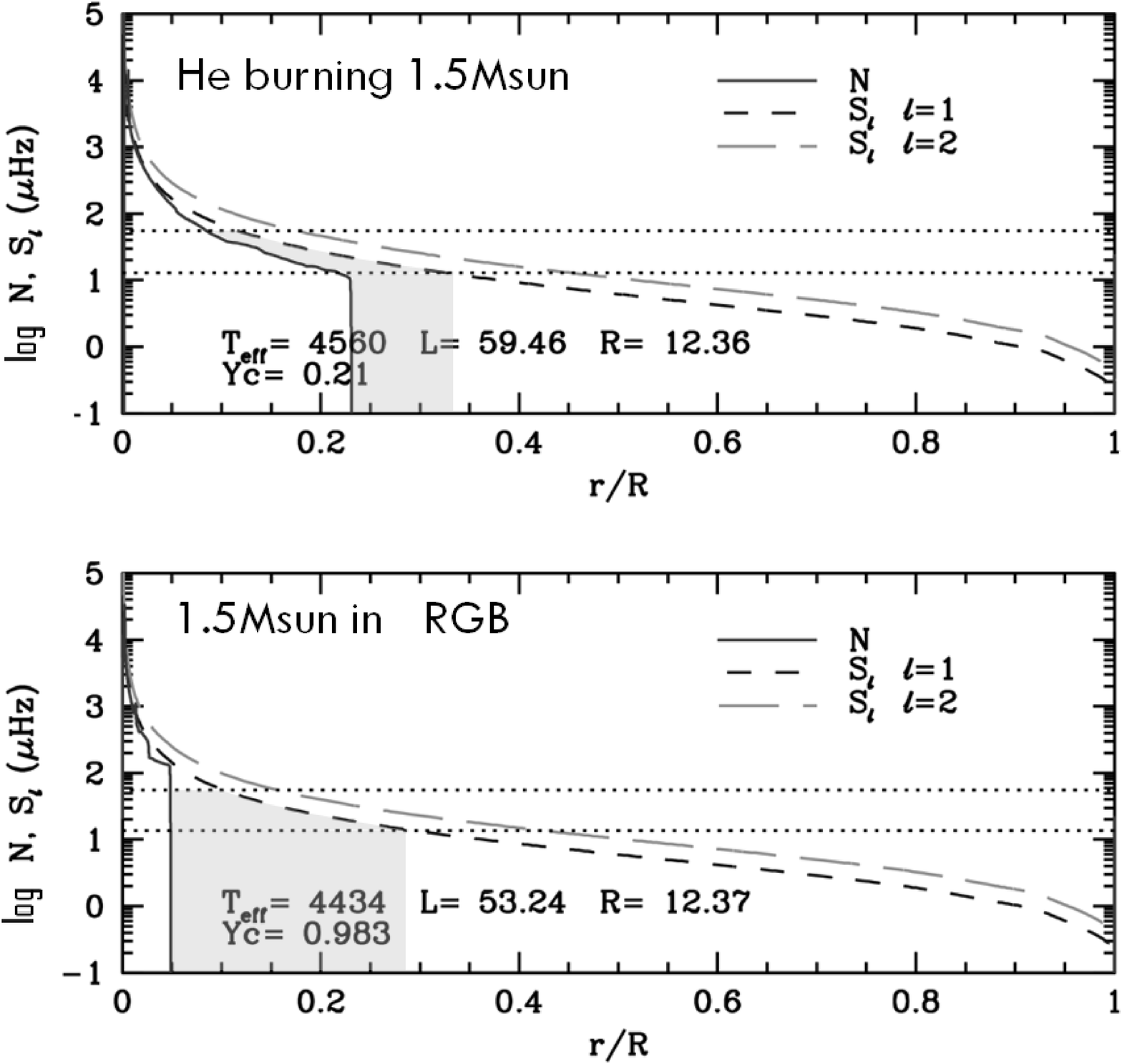}\hfill
\begin{minipage}[l]{0.3\linewidth}
\vspace*{-7.5cm}
\caption{Propagation diagrams for a 1.5~\msun\ star in two different evolutionary phases. Upper panel: central He burning phase, with central He mass fraction Yc=0.21, and lower panel,  RGB phase, both with the same stellar radius (R=12.35~\rsun). Horizontal lines indicate the solar-like frequency domain, and  shadowed regions highlight the potential barrier  that separates the acoustic and gravity cavities  for modes in that domain. Brunt-V\"ais\"al\"a and Lamb frequencies  for dipole and quadrupole modes as a function of the relative radius   correspond to solid, dashed and long-dashed lines respectively.}
\label{fig_DP}       
\end{minipage}\hfill

\end{figure}

Red giants are cool stars with an extended and diluted convective envelope surrounding a dense core, which makes their structure and therefore their pulsation properties very different from those of the Sun. As in solar-like stars, however, their convective envelope can stochastically excite oscillation modes. The properties of oscillation modes depend on the behavior of the Brunt-V\"ais\"al\"a ($N$) and Lamb ($S_{\ell}$) frequencies. In a first approximation we can describe the radial displacement due to the perturbation as: 
\begin{equation}
\frac{d^2\xi_r}{d\,r^2}=-K_s(r)\,\xi_r
\end{equation}
with $K_s(r)=\frac{\omega^2}{c_s^2}\left(\frac{N^2}{\omega^2}-1\right)\left(\frac{S_{\ell}^2}{\omega^2}-1\right)$, $\omega$ is the oscillation frequency and  $c_s$ the sound speed. From this expression we identify two propagation domains or cavities where the motion has an oscillatory character: $\omega^2 > N^2,S_{\ell}^2$  and  $\omega^2 < N^2,S_{\ell}^2$, while the regions where these conditions are not satisfied are evanescent regions with the amplitude of motion exponentially increasing or decreasing. In the limit cases in which $\omega^2 \gg  N^2,S_{\ell}^2$ or $\omega^2 \ll N^2,S_{\ell}^2$ we deal with pure acoustic (p modes) and pure  gravity modes (g modes) respectively. Because of the contraction of stellar central regions during the post-main sequence  evolution,  $N$ reaches huge values ($ > 10^4\mu$Hz) in the central regions of red giant models. As a consequence, the frequency of g modes \citep[$\omega_g \sim \int N/r {\rm d}r$, see][]{tassoul80} increases  with respect to main-sequence models. On the other hand, the drop of mean density resulting from the expansion of the hydrogen rich envelope makes the frequency of pressure modes ($\omega_p \sim \left(\int_0^R dr/c\right)^{-1}$)  decrease. All that leads to an oscillation spectrum for red giants where in addition to radial modes, one finds a large number of non-radial modes with mixed g-p properties, that is, for a given frequency $\omega$ the mode can propagate in the envelope ($\omega^2 > S^2,N^2$, see Fig.~\ref{fig_DP}) as a pressure mode and also in the central region ($\omega^2 < S^2,N^2$) as a g mode.  In Fig.~\ref{fig_mixmod} we plot the horizontal and radial components of the displacement corresponding to one of these g-p mixed modes propagating in a 1.4~\msun\ model with a radius of 5~\rsun.

The dominant p or g character of these non-radial modes depends on the coupling between gravity and acoustic cavities and may be estimated from the value of the normalized mode inertia \citep[$E$, see e.g.][ and references therein]{jcd04}: 

\begin{equation}
E=\frac{\int_V \rho |\mbox{\boldmath $ \xi $}|^2 dV}{M\,|\mbox{\boldmath $\xi$}|^2_{ph}},
\end{equation}

\noindent where $\rho$ is the local density, the integration is done over the volume $V$ of the star with total mass $M$, and $ph$ refers  to the value of displacement at the photosphere.  Modes trapped in high-density regions, such as g modes, have high values of $E$, while pure p modes such as the radial ones have the lowest $E$. Depending on the coupling between the two cavities, some non-radial modes  may be well trapped in the acoustic cavity  and behave as p modes presenting a mode inertia close to that of the radial modes, while modes with strong mixed g-p character have larger $E$. The modes well trapped in the central region have the maximum $E$ and behave as pure g modes.

\begin{figure}
\vspace*{1cm}
\hspace{.8cm}
\includegraphics[scale=0.65]{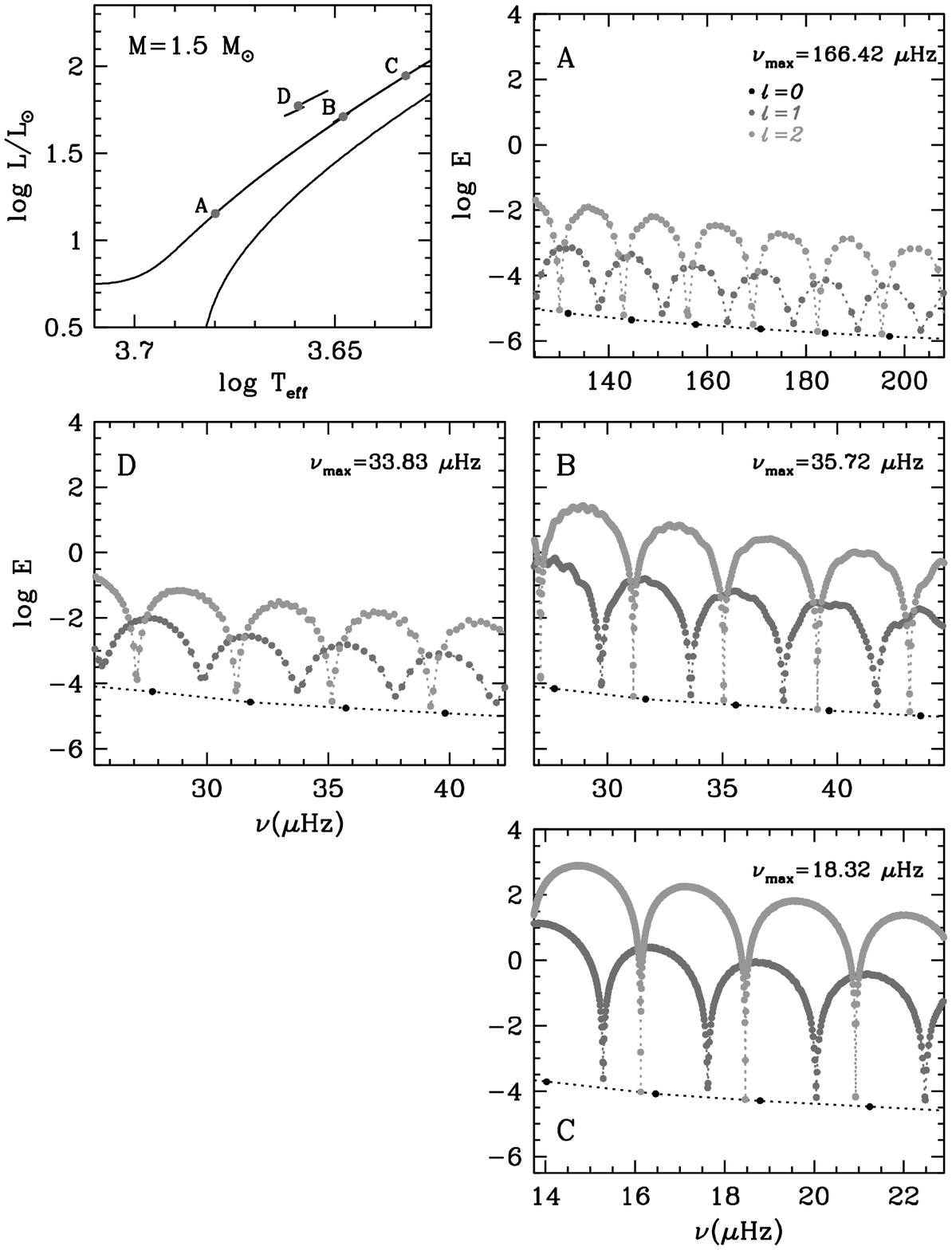}

\begin{minipage}[t]{0.45\linewidth}
\vspace*{-5cm}
\caption{Red giant evolutionary track for M=1.5~\msun, metal mass fraction Z=0.02 and initial helium mass fraction Y=0.278. Labels from A to D indicate the models whose spectra are shown in the corresponding panels. A to D panels: Inertia as a function of frequency  for radial (black), dipole (light gray) and quadrupole (dark grey) modes. $\nu_{\rm max}$ is the frequency at the maximum amplitude as expected from scaling laws \citep{KB95}.}
\label{fig_evol}       
\end{minipage}

\end{figure}

Figure~\ref{fig_DP} shows the  propagation diagrams  for dipole and quadrupole modes corresponding to a  1.5~\msun\ red giant on the red giant branch (upper panel) and in the  central He burning phase (lower panel). The horizontal lines indicate the frequency domain where we expect solar-like oscillations \citep[see][]{KB95}. From these diagrams we observe that:  {\it i)} the evanescent region is smaller for dipole modes than for quadrupole ones; {\it ii)} the importance of the evanescent region ($\int K_s {\rm d}r$), or the potential barrier that separates the two cavities,  depends on the evolutionary state. As a consequence, the interaction between the two cavities is more important for $\ell=1$ modes, and their mixed character changes with the evolutionary state. The number of g modes by frequency interval  can be estimated from the asymptotic theory \citep{tassoul80} as:
\begin{equation}
n_g\propto (\ell\,(\ell+1))^{1/2}\int\frac{N}{r} dr.
\end{equation}
\noindent
So, the density of oscillation modes is larger for quadrupole modes and its value also changes with the evolutionary state.

 The amplitude of modes at different frequencies in the oscillation spectrum results from the balance between excitation and damping rates \citep[][ and  Dupret in these proceedings]{Dziembowski01,Dupret09}, nevertheless, an estimation of the relative amplitude of different modes can be provided by the normalized mode inertia \citep{Houdek99}. In this framework, the amplitude is inversely proportional to $E^{1/2}$ \citep[see e.g. ][ and references therein]{jcd04}.

Figure~\ref{fig_evol} (panels A to D) presents the inertia of the oscillation modes as a function of frequency for radial, dipole and quadrupole modes for models of red giant stars in different evolutionary states. The corresponding models are indicated in the HR diagram (Fig.~\ref{fig_evol} left-top panel). Radial modes have the lowest inertia, and between two consecutive radial modes there is large number of non-radial modes with an inertia that can vary by several orders of magnitude. Between two radial modes there is always an  $\ell=2$ mode  with an inertia close to that of the adjacent radial mode. The  inertia of dipole modes ($E_{\ell=1}$)  presents as well a  minimum between two consecutive radial modes but, depending on the evolutionary state, that minimum is not always well defined, and the difference between the  value of $E_{\ell=1}$ and that corresponding to the radial mode varies. The central density of the  He burning (He-B hereafter) model is  10 times smaller than that of a  RGB model with the same radius and mass. Consequently, in the RGB phase, the high potential barrier between the acoustic and the gravity cavities reduces  the interaction between p and g modes, and dipole modes  with $E_{\ell=1}\sim E_{\ell=0}$ behave as pure p modes and show a regular frequency spacing. For He-B models,  the coupling between these cavities is more important and $\ell=1$ modes are mixed modes with $E_{\ell=1}> E_{\ell=0}$. Nevertheless, $E_{\ell=1}$ still presents a minimum and even if $E_{\ell=1}> E_{\ell=0}$,  the modes can be still considered, based on the value of $E$, as observable modes. 
If $E$ is a good proxy of the amplitudes, the results presented in Fig.~\ref{fig_evol} suggest that, contrarily to pure p-mode oscillation spectra of main sequence solar-like pulsators,  the number of non-radial modes  potentially observable in the frequency interval between two consecutive radial modes depends on $\ell$, on the evolutionary state, and on the precision of the time series, both in sensitivity and in frequency resolution.

\section{Diagnostic power of red giant oscillation spectra}

In the framework of the asymptotic theory for p modes \citep{vandakurov67, tassoul80, gough86} the frequencies of two modes of same degree and consecutive order are separated by a constant value $\langle \Delta\nu\rangle$ which is approximately independent of $\ell$ for low degree modes. The asymptotic theory is no longer valid for mixed modes or  in regions with rapid varying physical quantities, nevertheless, the modes partially or well trapped in the acoustic cavity (hence  with a dominant p character) show such a regular pattern. These modes have been  used to compute the large and small frequency separations and to analyse their behavior with stellar parameters and evolutionary state, both in recent  theoretical \citep{MontalbanAN, MontalbanApJL} and observational works based on CoRoT and \Kepler\ satellite data \citep{Beddingetal10, Huber10, MosserRedG, Carrier10}.

In the domain of validity of the asymptotic approximation for pressure modes, the spectrum of p modes can be described by \citep[e.g.][]{tassoul80}:
\begin{figure}[b]
\hspace{0.2cm}
\includegraphics[scale=.375]{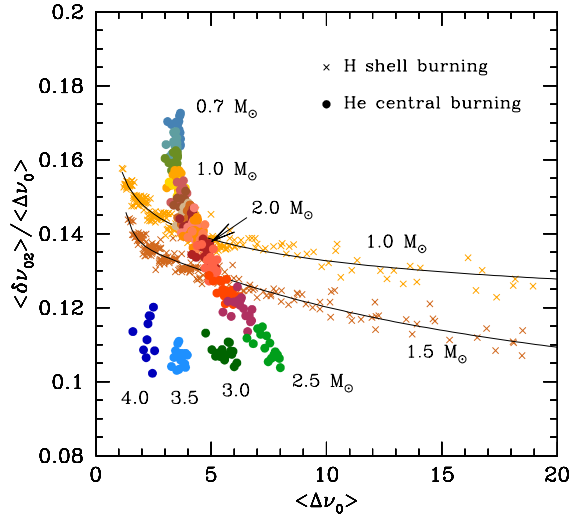}\hfill
\begin{minipage}[t]{0.34\linewidth}
\vspace*{-6.0cm}
\caption{Normalized small separation vs large separation of radial modes for models in the RGB phase (cross) with masses 1.0 and 1.5~\msun\,  chemical composition Z=0.006, 0.01, 0.02,0.03, Y=0.25 and 0.278, and three different treatments of convection. Dots correspond to models burning He in the center, with masses between 0.7 and 2.3~\msun ($\Delta M=0.1$) and between 2.5 and 4.0~\msun\ ($\Delta M=0.5$) and chemical composition Z=0.02, Y=0.278.}
\label{fig_d02LS}       
\end{minipage}\hfill
\end{figure}

\begin{equation}
\nu_{n\ell}=(n+\frac{\ell}{2}+\epsilon)\,\Delta\nu - \frac{\ell\,(\ell+1)}{4\,\ell+6}\delta \nu_{\ell}
\end{equation}

\noindent where $\Delta\nu=\nu_{n\,\ell}-\nu_{n-1\,\ell}$ is  the inverse of twice the sound travel time between the surface and the center, and is thus proportional to the square root of the mean density of the star. $\delta\nu_{n\ell}=\nu_{n\,\ell}-\nu_{n-1\,l+2}$ on the other hand, reflects the behaviour of the sound speed  mostly in the central regions, and hence is linked to the central density and to the  stellar evolutionary state. The dependence of $\delta\nu$ and $\Delta\nu$ on stellar mass and evolution is usually adopted as a good seismic diagnostic allowing to derive, if the chemical composition is known, the stellar mass and age for main sequence solar-like pulsators \citep[see e.g.][]{jcd88}. For red giants, however, $\delta\nu_{02}$  does not seem to provide much more information than $\Delta\nu$. Predictions from RGB models suggest  a linear dependence on $\Delta\nu$, with a slope that slightly increases as the mass decreases. Chemical composition and convection treatment do not strongly affect this linear dependence that is mainly dominated by the mass and radius of the star \citep{MontalbanAN,MontalbanApJL}. The observational results obtained from the first 34 days of {\it Kepler} operations that covered stars with $\Delta\nu > 8\,\mu$Hz were well fitted by the linear relation  $\delta_{02}\sim 0.125\Delta\nu$ \citep{Beddingetal10}. The comparison with theoretical models suggests that the observed sample is dominated by RGB stars with masses around 1.3~\msun. The extension of these observations to 134d allowed us to  consider stars  with  $\Delta\nu$ as low as  $~2\,\mu$Hz and to show that the slope of the relation and the scatter increase for $\Delta\nu <6\,\mu$Hz \citep{Huber10}. In fact, at $\Delta\nu\sim\,4\mu$Hz the red giant population is dominated by red clump stars \citep{MiglioPop09} that, as shown in Fig.~\ref{fig_d02LS}, follow a different trend. The comparison between this figure and  Fig.~10 in \cite{Huber10} indicates that predictions from theory are in good agreement with the observed behavior of $\delta\nu_{02}$  in  RGB and red clump stars.
 
For main sequence stars,  the frequency separation $\delta\nu_{01}=(\nu_{n,0}-2\,\nu_{n1}+\nu_{n+1,0})/2$  has also  been suggested as a good indicator of the evolutionary state \citep[e.g.][]{RoxburghVorontsov03}, and in the framework of  asymptotic approximation of p modes, $\delta\nu_{01} = 1/3 \delta\nu_{02}$. This relation is neither matched by theoretical  models of red giants nor by the observational data obtained with {\it Kepler} and CoRoT.  Nevertheless, the value of $\delta\nu_{01}$ may contain relevant information. In fact the sample of red giants studied in \cite{Beddingetal10}, that corresponds to low luminosity RGB stars ($L<30L_{\odot}$) and the CoRoT sismo target HR7349 \citep{Carrier10} show  values of  $\delta\nu_{01}$ close to zero or negative. Theoretical studies \citep{MontalbanAN,MontalbanApJL} on the other hand, predict that  models in the RGB present values of $\delta\nu_{01}$ close to zero or negative, while more massive stars in the He-B phase have positive values. Moreover, those studies show that the value of $\delta\nu_{01}$  seems correlated with the distance between the $\ell$=1 turning point and the bottom of the convective envelope, thus, models with the turning point inside the convective envelope present negative or very small value of  $\delta\nu_{01}$.

\subsection{Deviation from asymptotic approximation: Evolutionary state}

The mean value of the large frequency separation decreases as the star ascends the RGB with a denser and denser core and a more and more diffuse envelope. On the basis of  $\langle \Delta\nu\rangle$ alone, however, it is not possible to distinguish  among different evolutionary states, i.e. ascending RGB, descending RGB, or core-He burning. From the asymptotic theory  $\Delta\nu$ should be  constant  and $\ell$-independent, however,  we are out of its  validity domain and  $\Delta\nu$ varies with frequency and with $\ell$. A measurement of how far we are from the asymptotic behaviour could be the scatter of $\Delta\nu$ values ($\sigma (\Delta\nu_\ell)$) with respect to its average value in the domain of frequency where solar-like oscillations are expected.
The computation of these quantities for models with stellar  masses between 1 and 5 \msun\ from the bottom of the RGB until the exhaustion of He in the center shows that $\sigma (\Delta\nu_\ell)$ depends on $\ell$: while its  value is very small for $\ell=0$ and 2, it  is generally larger  for $\ell=1$ and its value depends strongly on the evolutionary state: {\it i)} the scatter in the $\ell=1$ ridge in an \'echelle diagram representation of the oscillation spectra \citep[e.g.][]{grec83} decreases as luminosity ($\Delta\nu$) increases (decreases) during the RGB phase, since  the increase of density contrast leads to decoupling the acoustic and gravity cavities and modes more likely observed behave as pure p modes. {\it ii)} For models in the He-B phase, the central density decreases leading to a spectrum of dipole modes dominated by p-g mixed modes. As a consequence, the width of the $\ell=1$ ridge will be more larger for those models than for the RGB ones at the same luminosity \citep{MontalbanAN}.
This result is very important because it suggests that the aspect of the $\ell=1$ oscillation spectrum is able to reveal the evolutionary state of red giant stars otherwise very close un their global properties \citep{MontalbanAN,MontalbanApJL}.

From the observational point of view, the population of red giants studied by \cite{Beddingetal10} is dominated by stars at the bottom of the RGB with masses between 1. and 1.5~\msun and showed a folded \'echelle diagram with  a larger dispersion in the ridge corresponding to dipole modes than in those for radial and quadrupole ones. On the other hand, the sample of red giants observed by CoRoT during the first two runs is dominated by red clump stars \citep{MiglioPop09, MosserRedG} and the analysis of their oscillation properties by \cite{MosserUniversal} showed that all the modes are arranged  in almost vertical lines corresponding to different radial orders, and that a large dispersion is found for dipole modes  at $\Delta\nu\sim 4\,\mu$Hz, that is the value of the large separation corresponding to the red-clump luminosity. The comparison between these observational results and the theoretical predictions for the same stellar population \citep{MontalbanApJL} is noteworthy.

As can be seen in Fig.~\ref{fig_evol},  around the minima of inertia, there may be several dipole modes with close values of the inertia and, therefore, with expected similar  amplitudes. As the observation time of red giants has increased, the frequency resolution has been noticeably  improved  and it has been possible not only to detect  in the spectra of red giant stars the acoustic modes, but also to distinguish around those a forest of  mixed modes. The frequency (or period) separation  between these modes around to the minimum of inertia, also depends on the evolution state and the mass of the model. Models with a large density contrast show a frequency difference between consecutive modes that is much smaller than for models at the bottom of the RGB or in the He-B phase. These properties have been measured by \cite{Bedding11Nat} in the spectra of red giants observed by \Kepler\  and also in those of CoRoT red giants \citep{MosserMixmods}. At a given $\Delta\nu$ (that of the red clump) the difference of period ($\Delta P$) between consecutive modes gathers the stars in two groups: one characterized by targets with  $\Delta P > 100$~s  and one with $\Delta P < 60$~s. The comparison with theoretical computations allows us to identify  these two groups with stars that are burning He at the center, for the first group, and with stars that are still burning H in a shell during the ascending RGB for the second one, and then  to lift the degeneracy between RGB and He-B models with the same $\Delta\nu$ and $\nu_{\rm max}$.

\subsection{ Deviation from asymptotic approximation: Glitch associated to HeII ionization region}
In the asymptotic approximation it is assumed that the physical quantities describing the equilibrium structure of the star present variations with a characteristic scale larger than the wavelength of the mode. Any localized region of sharp variation of the sound speed (so-called acoustic glitches)   induce in the frequencies an oscillatory component with a periodicity related to the sound-travel time measured from that region to the surface of the star (acoustic depth) \citep{Vorontsov88,Gough90}.  The amplitude of this oscillatory component depends on the sharpness of the glitch and decreases with frequency because, as $\nu$ increases,  the wavelength of the mode becomes comparable with or less than the extent of the glitch. These sharp variations of the structure are found in regions of rapidly changing chemical composition, in ionization zones of major chemical elements, or in regions where the energy transport switches from radiative to convective. The analysis of periodic variations in the frequencies, $\Delta\nu$, or second frequency differences, allowed the derivation of the depth of the solar convective envelope and its He content with a hight  precision \citep[see][for a review]{jcd02}.  In the case of red giants, the acoustic depth of the boundary of the convective envelope is very high and therefore the modulation of the frequency is not easily seen in the solar-like frequency domain. On the other hand,  the variation of the adiabatic index $\Gamma_1$ due to the second ionization of He is located at $\sim 0.5$ of the total acoustic radius and its signal is clearly separated from that of the bottom of the convective zone making the extraction of its signal much more easy than for solar-like main sequence pulsators. The first detection and characterization of the  HeII signal for a red giant was obtained for  the CoRoT sismo target HR7349 \citep{MiglioHeII}, but this signal has also been detected in numerous exofield CoRoT red giants \citep{MosserRedG} and it is expected that the red giants that {\it Kepler} will observe during three years will provide very precise characterization of their HeII signal.  The relevance of the detection and characterization of the acoustic glitch associated with the HeII region in red giant stars is enormous: while global features of the spectra such as $\Delta\nu$ and $\nu_{\rm max}$  will provide us with the mass and radius of the star,  the HeII feature will allow potentially to derive the abundance of He in its  convective envelope, allowing finally to answer questions about  simple or multiple stellar populations to explain stellar cluster morphology, and to eliminate one of the main sources of degeneracy in the study of evolved stars (see  Gratton,  in  this volume).


\section{Conclusions}
We have summarized here some aspects of the enormous potential of solar like oscillations in red giants. These stars, intrinsically highly luminous,  take up a small range of colors  in the HR diagram  for a large domain of masses, chemical composition and evolutionary state. A simple and adiabatic analysis of its stochastically excited oscillations allow us to derive their fundamental parameters such as mass and radius (see Miglio in this volume), but also their evolutionary state, and potentially the abundance of He in their convective envelope. 
The agreement between observation and theoretical prediction is up to now remarkable, allowing to explain observations in the framework of standard stellar models. In a near future, when individual frequencies will  be extracted, their study will allow us to go deeper into the study of stellar structure of red giants and their precursors. It is worthwhile to stress here that with only the global features of these spectra and a simple analysis we have already taken a giant step in the study of stellar evolution.

%

%


\end{document}